%% file: ms.tex
\documentclass[12pt,preprint]{aastex}

\slugcomment{To appear in ApJ}

\begin{document}


\title{VERITAS observations of the BL Lac object 1ES 1218+304}






\author{
V.A. Acciari\altaffilmark{20,1},
E. Aliu\altaffilmark{23},
T. Arlen\altaffilmark{17},
M. Beilicke\altaffilmark{2},
W. Benbow\altaffilmark{1},
S.M. Bradbury\altaffilmark{4},
J.H. Buckley\altaffilmark{2},
V. Bugaev\altaffilmark{2},
Y. Butt\altaffilmark{24},
K.L. Byrum\altaffilmark{5},
O. Celik\altaffilmark{6}, 
A. Cesarini\altaffilmark{1,21},
L. Ciupik\altaffilmark{7},
Y.C.K. Chow\altaffilmark{6},
P. Cogan\altaffilmark{12},
P. Colin\altaffilmark{11},
W. Cui\altaffilmark{8},
M.K. Daniel\altaffilmark{4,\dagger},
T. Ergin\altaffilmark{3},
A.D. Falcone\altaffilmark{22},
S.J. Fegan\altaffilmark{6},
J.P. Finley\altaffilmark{8},
P. Fortin\altaffilmark{14,*},
L.F. Fortson\altaffilmark{7},
A. Furniss\altaffilmark{17},
G.H. Gillanders\altaffilmark{21},
J. Grube\altaffilmark{13},
R. Guenette\altaffilmark{12},
G. Gyuk\altaffilmark{7},
D. Hanna\altaffilmark{12},
E. Hays\altaffilmark{5,1},
J. Holder\altaffilmark{23},
D. Horan\altaffilmark{5},
C.M. Hui\altaffilmark{11},
T.B. Humensky\altaffilmark{10},
A. Imran\altaffilmark{9},
P. Kaaret\altaffilmark{18},
N. Karlsson\altaffilmark{7},
M. Kertzman\altaffilmark{15},
D.B. Kieda\altaffilmark{11},
J. Kildea\altaffilmark{1},
A. Konopelko\altaffilmark{8},
H. Krawczynski\altaffilmark{2},
F. Krennrich\altaffilmark{9},
M.J. Lang\altaffilmark{21},
S. LeBohec\altaffilmark{11},
G. Maier\altaffilmark{12},
A. McCann\altaffilmark{12},
M. McCutcheon\altaffilmark{12},
P. Moriarty\altaffilmark{20},
R. Mukherjee\altaffilmark{14},
T. Nagai\altaffilmark{9},
J. Niemiec\altaffilmark{9,\ddagger},
R.A. Ong\altaffilmark{6},
D. Pandel\altaffilmark{18},
J.S. Perkins\altaffilmark{1},
M. Pohl\altaffilmark{9},
J. Quinn\altaffilmark{13},
K. Ragan\altaffilmark{12},
L.C. Reyes\altaffilmark{10},
P.T. Reynolds\altaffilmark{19},
H.J. Rose\altaffilmark{4},
M. Schroedter\altaffilmark{9},
G.H. Sembroski\altaffilmark{8},
A.W. Smith\altaffilmark{1,4},
D. Steele \altaffilmark{7},
S.P. Swordy\altaffilmark{10},
J.A. Toner\altaffilmark{1,21},
L. Valcarcel\altaffilmark{12},
V.V. Vassiliev\altaffilmark{6},
R. Wagner\altaffilmark{5},
S.P. Wakely\altaffilmark{10},
J.E. Ward\altaffilmark{13},
T.C. Weekes\altaffilmark{1},
A. Weinstein\altaffilmark{6},
R.J. White\altaffilmark{4},
D.A. Williams\altaffilmark{17},
S.A. Wissel\altaffilmark{10},
M. Wood\altaffilmark{6},
B. Zitzer\altaffilmark{8}
}


\altaffiltext{1}{Fred Lawrence Whipple Observatory, Harvard-Smithsonian 
Center for Astrophysics, Amado, AZ 85645, USA}

\altaffiltext{2}{Department of Physics, Washington University, St. 
Louis, MO 63130, USA}

\altaffiltext{3}{Department of Physics, University of Massachusetts, 
Amherst, MA 01003-4525, USA}

\altaffiltext{4}{School of Physics and Astronomy, University of Leeds, 
Leeds LS2 9JT, UK}

\altaffiltext{5}{Argonne National Laboratory, 9700 S. Cass Avenue, 
Argonne, IL 60439, USA}

\altaffiltext{6}{Department of Physics and Astronomy, University of 
California, Los Angeles, CA 90095, USA}

\altaffiltext{7}{Astronomy Department, Adler Planetarium and Astronomy 
Museum, Chicago, IL 60605, USA}

\altaffiltext{8}{Department of Physics, Purdue University, West 
Lafayette, IN 47907, USA}

\altaffiltext{9}{Department of Physics and Astronomy, Iowa State 
University, Ames, IA 50011, USA}

\altaffiltext{10}{Enrico Fermi Institute, University of Chicago, 
Chicago, IL 60637, USA}

\altaffiltext{11}{Physics Department, University of Utah, Salt Lake 
City, UT 84112, USA}

\altaffiltext{12}{Physics Department, McGill University, Montreal, QC 
H3A 2T8, Canada}

\altaffiltext{13}{School of Physics, University College Dublin, 
Belfield, Dublin, Ireland }

\altaffiltext{14}{Department of Physics and Astronomy, Barnard College, 
Columbia University, NY 10027, USA}


\altaffiltext{16}{Department of Physics, Grinnell College, Grinnell, IA 
50112-1690, USA}

\altaffiltext{17}{Santa Cruz Institute for Particle Physics and 
Department of Physics, University of California, Santa Cruz, CA 95064, 
USA}

\altaffiltext{18}{Department of Physics and Astronomy, University of 
Iowa, Van Allen Hall, Iowa City, IA 52242, USA}

\altaffiltext{19}{Department of Applied Physics and Instrumentation, 
Cork Institute of Technology, Bishopstown, Cork, Ireland}

\altaffiltext{20}{Department of Life and Physical Sciences, Galway-Mayo 
Institute of Technology, Dublin Road, Galway, Ireland}

\altaffiltext{21}{Physics Department, National University of Ireland, 
Galway, Ireland}

\altaffiltext{22}{Department of Astronomy and Astrophysics, Penn State 
University, University Park, PA 16802, USA}

\altaffiltext{23}{Department of Physics and Astronomy, Bartol Research 
Institute, University of Delaware, Newark, DE 19716, USA}

\altaffiltext{24}{Smithsonian Astrophysical Observatory, Cambridge, MA 
02138, USA}


\altaffiltext{26}{Max-Planck Institut for Extraterrestrial Physics 
(MPE), Giessenbachstrasse, 85748 Garching, Germany}

\altaffiltext{$\dagger$}{Now at: Department of Physics, Durham 
University, South Road, Durham, DH1 3LE, U.K.} 

\altaffiltext{$\ddagger$}{Now at: Instytut Fizyki J\c{a}drowej PAN, ul.  
Radzikowskiego 152, 31-342 Krak\'ow, Poland} 

\altaffiltext{*}{Corresponding author: fortin@phys.columbia.edu}


\begin{abstract}
The VERITAS collaboration reports the detection of very-high-energy (VHE) gamma-ray emission from the high-frequency-peaked BL Lac object 1ES 1218+304 located at a redshift of $z=0.182$. A gamma-ray signal was detected with a statistical significance of 10.4 standard deviations (10.4 $ \sigma$) for the observations taken during the first three months of 2007, confirming the discovery of this object made by the MAGIC collaboration. The photon spectrum between $\sim160$ GeV and $\sim1.8$ TeV is well described by a power law with an index of $\Gamma = 3.08 \pm 0.34_{stat} \pm 0.2_{sys}$. The integral flux is $\Phi(\textrm{E} > 200 \textrm{ GeV}) = (12.2 \pm 2.6) \times 10^{-12} \mathrm{cm^{-2}s^{-1}}$, which corresponds to $\sim6$\% of that of the Crab Nebula. The light curve does not show any evidence for VHE flux variability. Using lower limits on the density of the extragalactic background light in the near to mid-infrared we are able to limit the range of intrinsic energy spectra for 1ES~1218+304. We show that the intrinsic photon spectrum has an index that is harder than $\Gamma = 2.32 \pm 0.37_{stat}$. When including constraints from the spectra of 1ES~1101-232 and 1ES~0229+200, the spectrum of 1ES~1218+304 is likely to be harder than $\Gamma = 1.86 \pm 0.37_{stat}$.

\end{abstract}





\section{Introduction}

One of the major discoveries of EGRET on the Compton Gamma Ray Observatory was the detection of high-energy emission from more than 60 active galactic nuclei (AGN) of the blazar class \citep{hartman}. Blazars, which include BL Lac objects and flat-spectrum radio quasars (FSRQs), are characterized by non-thermal emission and their spectral energy distributions (SED) contain two broad peaks. The low-energy peak (radio to UV or X-rays) is commonly interpreted as synchrotron radiation from ultra-relativistic electrons moving along a plasma jet pointing towards the observer. The origin of the second peak is less certain. Several models, from pure leptonic or hadronic models to leptonic/hadronic hybrid models, can explain the high-energy peak (X-rays to TeV gamma-rays) \citep[see][and references therein]{Bottcher:2007fr}.

In the TeV energy range, 19 blazars and one radio galaxy (M 87) have been established as emitters of TeV gamma-rays.
High-frequency-peaked BL Lac (HBL) objects \citep{Padovani:1995p2580} are a subclass of blazars characterized by a synchrotron peak at X-ray energies, unlike quasars that have higher luminosity and a synchrotron peak at optical/infrared energies.
With the exception of the recently discovered low-frequency-peaked BL Lac (LBL) object BL Lacertae \citep{Albert:2007mz}, the intermediate-frequency-peaked BL Lac (IBL) object W Comae \citep{Swordy:2008}, and the flat spectrum radio quasar 3C 279 \citep{Albert:2008p3043}, all TeV blazars are high-frequency-peaked BL Lac objects.
Some of these objects can show rapid, down to a few minutes,  flux variability at TeV energies \citep[see e.g.,][]{Gaidos:1996p3096,Aharonian:2007p762,Albert:2007p3044}.

Broadband observations from the radio to very-high-energy (VHE) gamma-rays are necessary to understand the physics of the jets and emission mechanisms. VHE observations can also help constrain the intensity and spectrum of the extragalactic background light (EBL) \citep{Stecker:1992p2571}, which are important parameters for cosmologists to test our understanding of structure and star formation in the Universe. Produced by stars and partially reprocessed by dust, the EBL at near to mid-infrared (near-IR, mid-IR) wavelengths is a strong absorber of TeV gamma-rays via pair production \citep{Gould:1967p2336}. Direct measurements of the EBL in this range are particularly difficult due to the dominant foreground of zodiacal light. The measurement of high-quality energy spectra for blazars over the energy range from 100 GeV to 10 TeV can be used to gain information about the EBL \citep{Dwek:2005ve}. The analysis of TeV gamma-ray energy spectra of several blazars suggests that the intensity of the EBL in the near-IR to mid-IR band is close to the lower limit from source counts measured by the Hubble and Spitzer space telescopes \citep[see][]{Aharonian:2006p1822,Aharonian:2007p1477,Mazin:2007p2248}.

Based on SED modeling and \textit{Beppo}SAX X-ray spectra, several of the HBLs were predicted to be TeV sources and several of them have indeed been detected at TeV energies \citep{Costamante:2002rt}.
1ES~1218+304 is an X-ray-bright ($F_{1 keV}>2\mu $Jy) HBL object located at a redshift $z=0.182$ \citep{Bade:1998ly}, and it  was predicted to be a good TeV candidate based on the position of its synchrotron peak at high energy and sufficient radio-to-optical flux. It was first detected at very high energies by the MAGIC telescope \citep{Albert:2006yq}. This object was also the target for a brief HESS observation campaign in May 2006 \citep{Aharonian:2008p1531}. An upper limit on the integral flux above 1 TeV corresponding to $\sim6$ times the flux determined from an extrapolation of the MAGIC spectrum was reported. 

In this paper we report on the detection of 1ES 1218+304 in VHE gamma rays with VERITAS. A description of the telescopes and a summary of VERITAS observations of 1ES 1218+304 are presented in \S 2. The results of the data analysis are presented in \S 3, and a discussion of our results in the context of the EBL density is presented in \S 4.

\section{Observations}

The VERITAS observatory consists of an array of four 12-meter diameter imaging atmospheric Cherenkov telescopes (IACTs) located at the Fred Lawrence Whipple Observatory ($31^\circ 40'30''$ N, $110^\circ57'07''$ W, 1268 m a.s.l) in southern Arizona \citep{T.-C.-Weekes:2002lr}. The telescopes use the Davies-Cotton design \citep{Davies:Cotton} and each utilize 350 front-aluminized and anodized hexagonal glass facets with a total mirror area of 106 m$^2$. Each camera consists of 499 photomultiplier tubes (PMTs) separated by $0.15^\circ$ and covers a $3.5^\circ$ field of view. Light concentrators reduce the dead space between PMTs and decrease the amount of ambient light seen by the PMTs. The analog signals from the PMTs are pre-amplified in the camera before being sent through coaxial cables to an electronics room located at the base of each telescope. VERITAS uses a three-level trigger system to reduce the rate of background events caused by fluctuations in the night sky light and by cosmic-ray showers while retaining multi-telescope images consistent with gamma-ray showers. The analog PMT signals are digitized using custom-designed 500 mega-sample-per-second flash-analog-to-digital converters (FADCs) and the data are archived to disk. Additional technical details and more in-depth descriptions of the performance of the telescope can be found in \citet{Holder:2006p2560} and \citet{Maier:2007kx}.

The first two telescopes were operated in the stereoscopic observation mode from March 2006 and the third and fourth telescopes came online in December 2006 and April 2007, respectively. VERITAS observed 1ES 1218+304 from January to March 2007 using three telescopes. The data were taken in \textit{wobble} mode where the source is offset from the center of the field of view by $0.5^\circ$ and the background is measured directly from different regions in the same field of view but away from the source region. After removing data taken under poor sky conditions or affected by various detector problems, we were left with a total observation time of 17.4 hours covering a range in zenith angle from $2^{\circ}$ to $35^{\circ}$, with an average zenith angle of $14^{\circ}$.

\section{Data Analysis \& Results}

The analysis of the data was performed using independent analysis packages \citep[see][for details on the analyses]{Daniel:2007lr}. All of these analyses yield consistent results.
After calculating the standard Hillas parameters \citep{hillas85}, images with an integrated charge less than 400 digital counts\footnote{An integrated charge of 400 digital counts calculated using a 10 ns integration window corresponds to $\sim$75 photoelectrons.} or with a distance from the center of the camera larger than $1.2^{\circ}$ were rejected.
The location of the shower direction in the field of view and the impact parameter of the shower core were calculated using stereoscopic techniques \citep{Hofmann:1999lr,Krawczynski:2006lr}. Events originating from a circular region with radius $\theta = 0.158^\circ$ centered on the position of 1ES 1218+304 are taken as the source region, and the background was estimated from the same field of view using the \textit{reflected-region} and \textit{ring background} methods, as described in \citet{Berge:2007p689}.  A set of \textit{scaled cuts} on the \textit{width} and \textit{length} parameters were used to identify gamma-ray events in the data \citep{Konopelko}. The cuts were optimized a priori with Monte Carlo simulations of gamma-ray and hadron induced air showers. The cuts used here are (-1.25 $<$ \textit{mean-scaled width/length} $<$ 0.5). An extensive set of Monte Carlo simulations was used to generate lookup tables to calculate the energy of the primary gamma rays. The effective area of the detector as a function of zenith angle and gamma-ray energy was also calculated from these simulations. The variability of the atmospheric conditions and the overall photon collection efficiency are the major contributors to the systematic error of the energy estimation.

\begin{figure}
\plotone{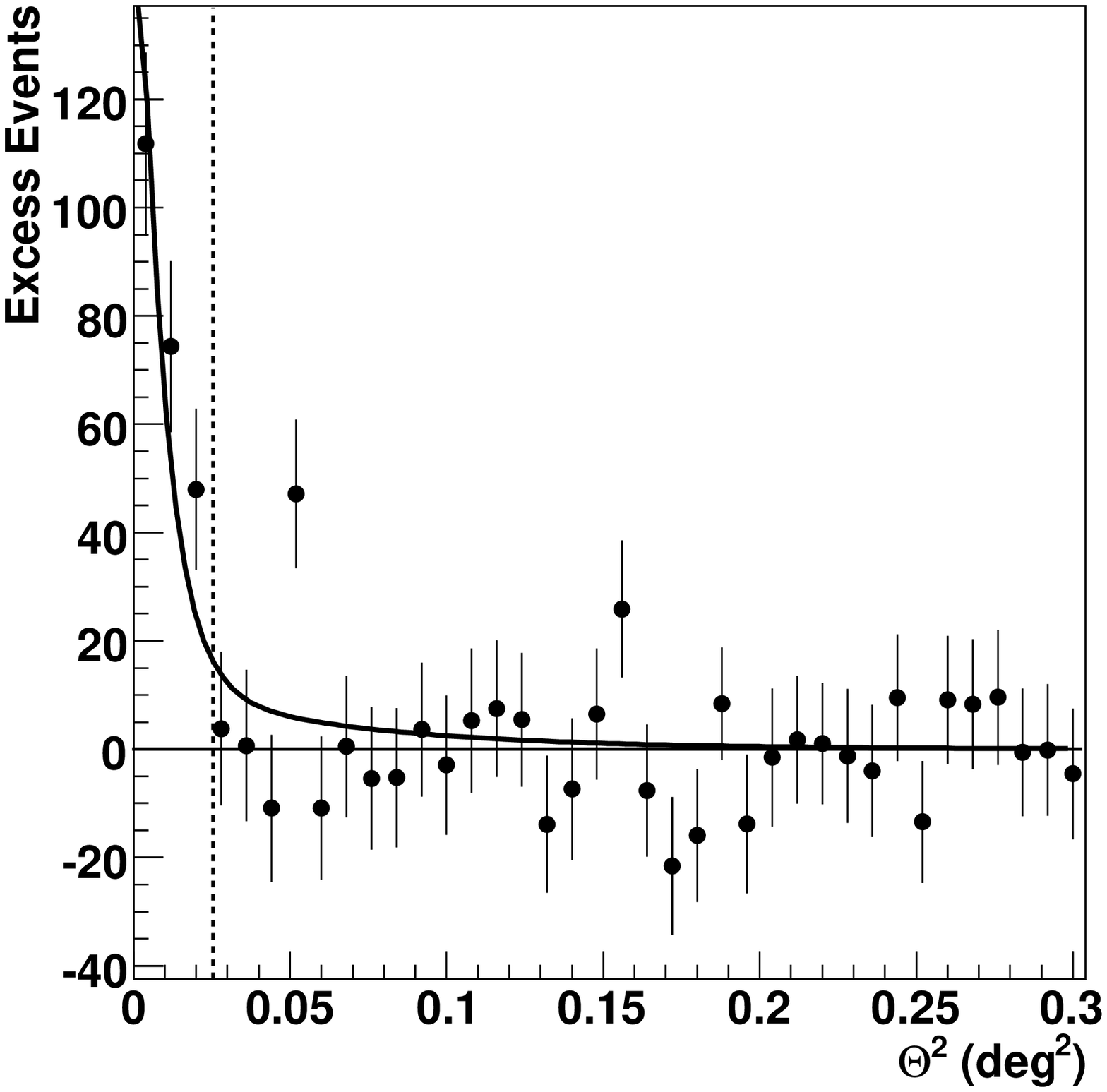}
\caption{Distribution of excess events as a function of the squared angular distance ($\Theta^2$) between the reconstructed shower directions and the nominal position of 1ES 1218+304. The dashed line indicates the boundary of the \textit{signal} region at $0.158^\circ$ (0.025 deg$^2$) as determined from optimization on the Crab Nebula. The solid curve shows the expectation for a point source as measured from data on the Crab Nebula.}\label{fig1}
\end{figure}

\begin{figure}
\plotone{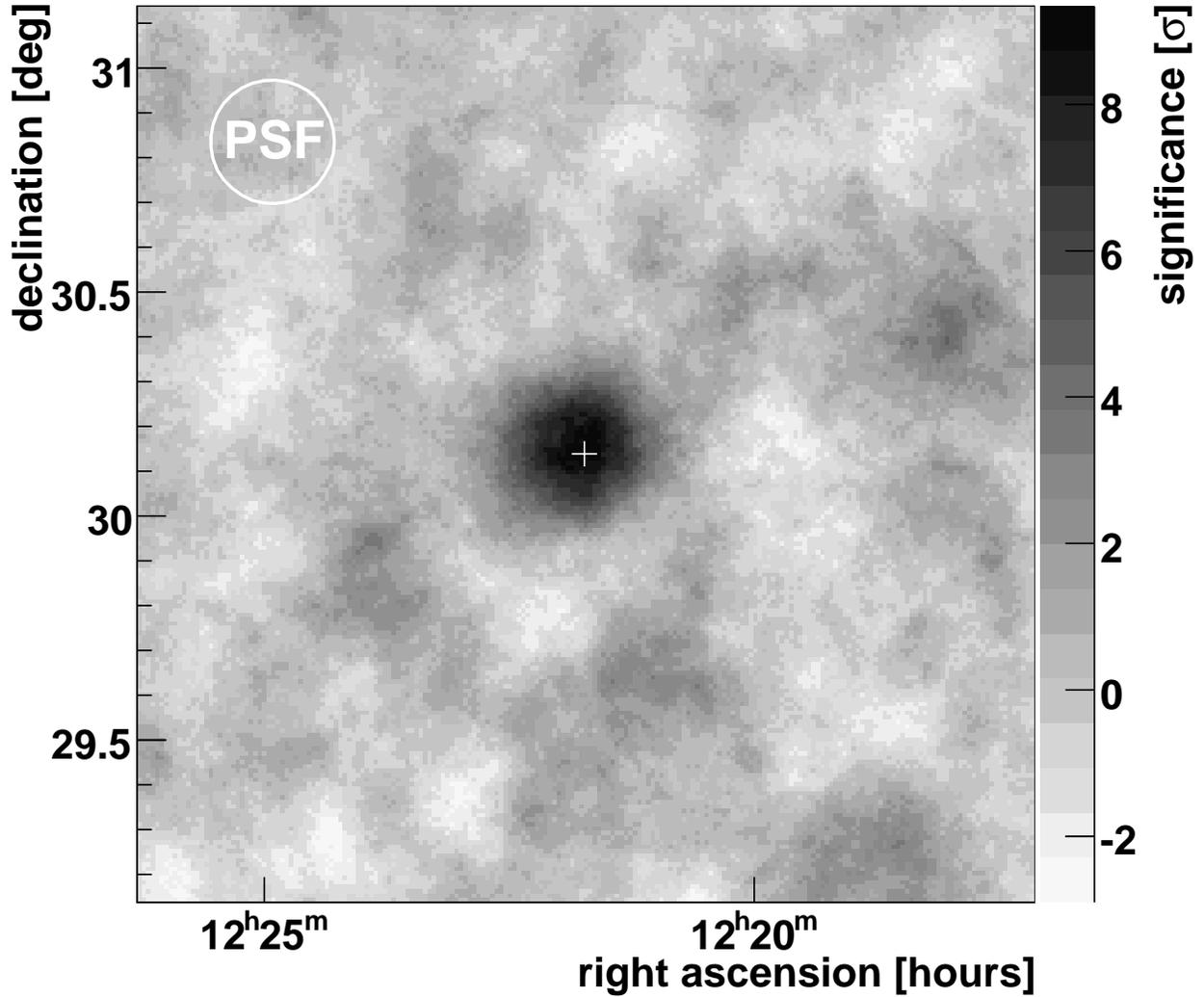}
\caption{Significance map of the region around 1ES 1218+304. The white cross indicates the position of the corresponding radio source. The white circle in the upper left corner shows the angular resolution of VERITAS (PSF). The ring background model was used to estimate the background and the significances were calculated using equation 17 in \citet{Li:1983lr}.} \label{fig2}
\end{figure}

Figure \ref{fig1} shows the number of excess events calculated using the \textit{reflected-region} method as a function of the squared angular distance ($\theta^2$) between the reconstructed shower directions and the nominal position of 1ES 1218+304. A clear excess is visible below the angular cut of $0.158^\circ$, corresponding to a statistical significance of 10.4 standard deviations using equation (17) in \citet{Li:1983lr} (617 signal events, 1466 background events with a normalization of  0.25). Figure \ref{fig2} shows a two-dimensional sky map of significances for a region centered on the radio coordinates of 1ES 1218+304 ($12^{h}21^{m}21.9^{s}, +30^{\circ}10'37''$ J2000) \citep{Becker:1995p3168}. The excess is compatible with that expected from a point source. Fitting a two-dimensional normal distribution to the uncorrelated excess map yields a peak position with coordinates ($12^{h}21^{m}26.3^{s}\pm2.8^{s}_{stat}\pm7.0^{s}_{sys}, +30^{\circ}11' 29''\pm33''_{stat}\pm1'30''_{sys}$ J2000), in good agreement with the radio coordinates.
The systematic uncertainty on the tracking of the VERITAS telescopes was $\pm90$ arcseconds at the time the data were taken on 1ES 1218+304\footnote{Pointing monitors were installed on all 4 telescopes during the summer of 2007. This is expected to improve the pointing accuracy significantly in the future.}.

\begin{figure}
\plotone{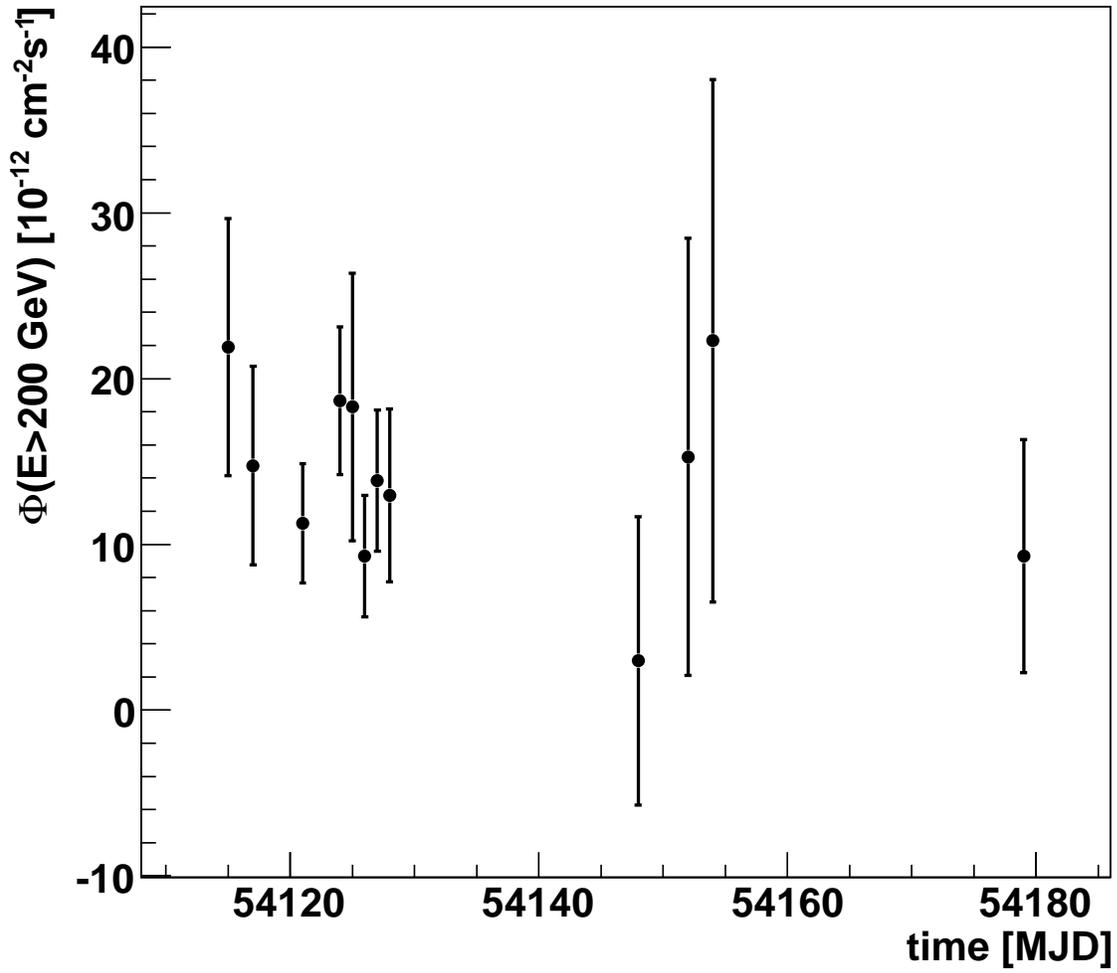}
\caption{Light curve of the integral photon flux above 200 GeV for the source 1ES 1218+304. Each point corresponds to the average daily flux and assumes a spectral shape of dN/dE $\propto$ E$^{-\Gamma}$ with $\Gamma = 3.08$. The error bars represent the statistical uncertainty.}\label{fig3}
\end{figure}

Figure \ref{fig3} shows the light curve of the integral flux above 200 GeV for the months of January, February and March 2007. The average integral flux was calculated for each day assuming a spectral shape of dN/dE $\propto$ E$^{-\Gamma}$ with $\Gamma = 3.08$. A statistical test ($\chi^2$/dof = 6.7/11) indicates that no statistically significant variability was detected in the data.

\begin{figure}
\plotone{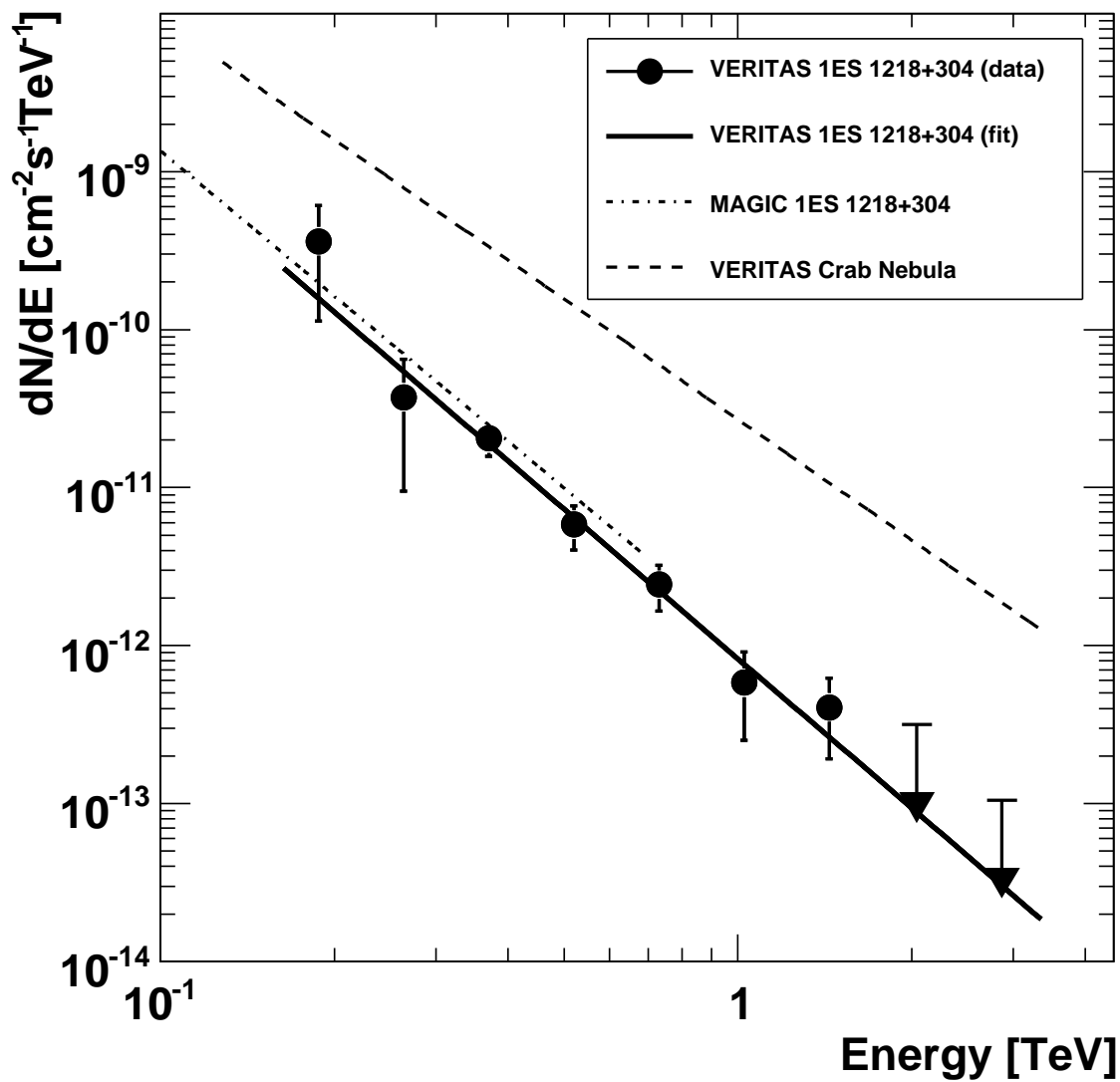}
\caption{Differential energy spectrum of VHE photons above 160 GeV for 1ES 1218+304. The markers indicate measured data points and the continuous line is a power-law fit. Downward pointing arrows correspond to upper flux limits (99\% probability, \citet{helene}) for bins with significances below two standard deviations. The dot-dash line shows the differential energy spectrum of 1ES 1218+304 measured by MAGIC \citep{Albert:2006yq} and the dash line shows the differential energy spectrum of the Crab Nebula measured by VERITAS (September to November 2006) for comparison.}
\label{fig4}
\end{figure}

Figure \ref{fig4} shows the time-average differential energy spectrum for gamma-ray energies between 160 GeV and 1.8 TeV. The shape is consistent with a power law ($\chi^{2}/\textrm{dof} = 2.1/5$) $\mathrm{dN/dE} = \mathrm{C\times(E/0.5TeV)}^{-\Gamma}$ with a photon index $\Gamma = 3.08 \pm 0.34_{stat} \pm 0.2_{sys}$ and a flux normalization constant $\mathrm{C} = (7.5 \pm 1.1_{stat} \pm 1.5_{sys}) \times 10^{-12} \mathrm{cm^{-2}s^{-1}TeV^{-1}}$. 
The integral flux is $\Phi(\textrm{E} > 200 \textrm{ GeV}) = (12.2 \pm 2.6) \times 10^{-12} \mathrm{cm^{-2}s^{-1}}$ which corresponds to $\sim6$\% of the flux of  the Crab Nebula above the same threshold.
The MAGIC collaboration measured the differential energy spectrum of 1ES 1218+304 between 87 GeV and 630 GeV during 6 nights of observations in 2005 January \citep{Albert:2006yq}. No evidence for time variability was detected. The VERITAS measurement extends the differential energy spectrum to 1.8 TeV and is statistically consistent with the results reported by the MAGIC collaboration.

\section{Discussion and conclusions}

Due to the relatively large redshift of 1ES 1218+304 ($z=0.182$) we expect significant attenuation from the interaction of high-energy gamma-rays with the low-energy photons of the EBL. TeV gamma-rays from extragalactic sources are expected to interact with optical low energy photons as they travel through inter-galactic space, leading to a cutoff in the measured gamma-ray spectrum. The optical depth $\tau$ of the attenuation is a complicated function of the gamma-ray photon energy, the distance to the source (redshift $z$), and the cross-section for pair production, and it is related to the density and spectral energy distribution of the cosmic background radiation. The physical importance of the EBL lies in its relation to galaxy formation and evolution and to the star formation history of the Universe. If the intrinsic spectra of blazars extend to $\sim 10$ TeV, absorption features in the blazar spectra can be used to learn about the EBL in the mid-IR band. TeV spectra of blazars therefore have the potential of providing independent constraints on the infrared background light density (see, e.g., \citet{Primack:2002p2585,Coppi:1999p2590}). In addition to inter-galactic absorption, the measured blazar spectra are shaped by intrinsic absorption in the blazar, as well as by Compton scattering in the Klein-Nishina regime \citep{Moderski:2005p2601}.

Intrinsic spectra of blazars are not known a priori. Nevertheless,  it is possible to use the EBL lower limits from galaxy counts to restrict the range of EBL scenarios/models and discern a corresponding range of  intrinsic blazar spectra. We present possible intrinsic energy spectra of 1ES~1218+304 that are compatible with EBL lower limits from galaxy counts.   Furthermore, we broaden our study and include the energy spectra of 1ES~1101-232 and 1ES~0229+200 as measured by H.E.S.S. \citep[see][respectively]{Aharonian:2006p1822,Aharonian:2007p1477} to reach a conclusion about the intrinsic energy spectra of this small sample of TeV blazars. The combination of energy spectra in the sub-TeV to TeV waveband at large redshifts (1ES~1101-232 and 1ES~1218+304 are two of the most distant HBLs) with energy spectra in the multi-TeV regime at a moderate redshift (1ES~0229+200) provides additional sensitivity to EBL spectra and the relative intensities in the near-IR and mid-IR. Consequently, this also leads to stronger constraints for the intrinsic spectra of these blazars. The combination of these three blazars allows us to make a statement about the hardness of blazar spectra in general, and sheds light on the question: how hard are intrinsic TeV blazar spectra? 

\begin{figure}
\epsscale{.70}
\plotone{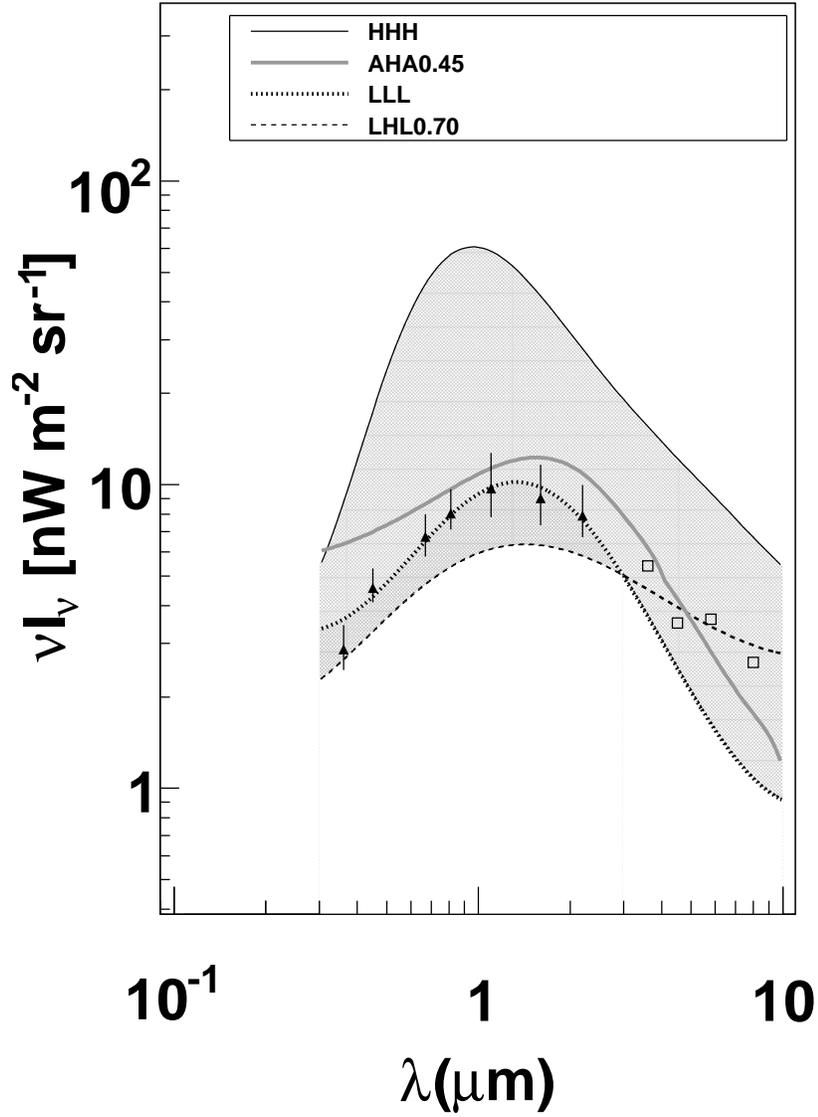}
\caption{The range of EBL densities that are considered for reconstructing the source spectrum
of 1ES 1218+304 is shown. The filled triangles represent lower limits from galaxy counts by 
\citep{Madau2000} whereas the open rectangles are lower limits from the Spitzer observatory
\citep{Fazio2004}.}
\label{fig5}
\end{figure}

In order to obtain the intrinsic spectrum of a blazar, the measured spectrum must be corrected by unfolding the effects of the EBL.  Figure \ref{fig5} shows the range of EBL scenarios  considered for unfolding the intrinsic spectrum of 1ES~1218+304. The lower limits from galaxy counts derived from the HST deep sky survey \citep{Madau2000} provide the lowest possible EBL in the optical to near-IR (filled triangles). Galaxy counts from the Spitzer infrared observatory \citep{Fazio2004} in the mid-IR are complementary (open quadrangles). 
A convenient parameterization of EBL scenarios, provided by \citet{Dwek:2005ve}, is used in this study for providing a limit to the hardness of the blazar spectrum of 1ES~1218+304 as it provides a wide range of EBL spectra with different near-IR to mid-IR ratios.
The EBL scenarios are parameterized using polynomials and are constrained by detections and lower limits in the UV, optical, and sub-millimeter wavebands. Further details can be found in \citet{Dwek:2005ve}, which presents 12 possible EBL realizations.
The optical depth was calculated using a flat universe cosmology with $H_{0}= 70~\rm km~s^{-1}$, $\Omega_{m} = 0.3$, and $\Omega_{\Lambda} = 0.7$.
For illustrative purposes we show two low EBL scenarios in Figure \ref{fig5} parameterized by \citet{Dwek:2005ve}.
The first, called LLL (dotted line) is an EBL realization representing a low-intensity stellar component, a low-intensity 15$\mu$m EBL flux, and a low-intensity of the far-IR flux, and falls significantly below the galaxy counts in the mid-IR. The second, called LHL0.70 (dashed line) is a scaled version of the LHL realization and falls below galaxy counts in the optical to near-IR. Furthermore, we have considered a large variety of EBL scenarios with different spectral indices and shapes that are within the boundaries of the shaded area in Figure \ref{fig5}. The upper bound is somehat arbitrary and was chosen to be the HHH (High near-IR, mid-IR and far-IR) scenario.  
  
Furthermore, for reference, we also derive the absorption-corrected intrinsic gamma-ray spectra for theoretical EBL models by  Stecker, Malkan, and Scully (2006, hereafter SMS).
In these models, the intergalactic IR photon flux and density were calculated using a {\sl backward-evolution} method which started with existing galaxy populations and modeled the luminosity evolution of these galaxies back in time.
It is, however, important to note that in this paper we do not adopt any particular model, we simply use constraints from EBL measurements to derive a lower limit to the hardness of the 1ES~1218+304 spectrum.

Table~1  shows the intrinsic energy spectra photon indices of 1ES~1218+304 for a range of EBL scenarios.
The ones that fall below the limit from galaxy counts (see Figure \ref{fig5}) are marked with an asterisk in Table~1 and are not viable, e.g., the LHL0.70, the LLL and the LLH scenarios.
Scenarios that are still compatible with the lower limits from galaxy counts represent the softest possible gamma-ray spectra.
When considering 1ES~1218+304 by itself, the softest intrinsic spectrum is described by a power law with $\rm dN/dE \propto E^{-2.32 \pm 0.37_{stat}}$ and is derived from a scaled version of the LHL scenario (LHL0.76).

\begin{figure}
\plotone{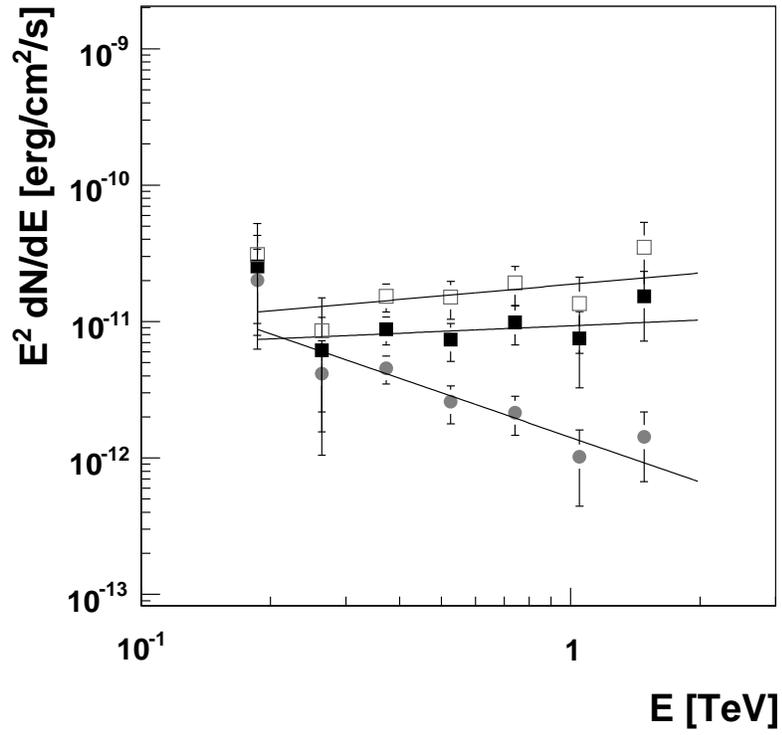}
\caption{Measured spectrum of 1ES1218+304 (circles) and de-absorbed spectrum for the AHA0.45  EBL scenario that produces the softest possible spectrum among all EBL scenarios considered (filled quadrangles). Furthermore we show the corrected spectrum for the SMS baseline model by \citet{Malkan:1998p2653,Stecker:2006p2224} (hollow quadrangles).}
\label{fig6}
\end{figure}

However, when applying this analysis to previously reported blazar spectra from 1ES~1101-232 and 1ES~0229+200, the LHL0.76 would require an extremely hard intrinsic spectrum for 1ES~0229+200\footnote{The data points for the energy spectra of 1ES~1101-232 and 1ES~0229+200 were provided to us by the H.E.S.S. collaboration. We have carefully checked the results of our power law fits against those published by the H.E.S.S. collaboration and they are in agreement. The correction for the EBL absorption was applied to the individual flux points and we fitted the absorption-corrected flux points with power laws.}.
Therefore, in order to provide a limit to the hardness of the blazar spectra based on these three sources we search for the softest possible intrinsic spectrum. As can be seen from Table~1, a search for the softest possible blazar spectrum among this sample of three  blazars yields an EBL scenario (AHA0.45) that still requires the spectrum of 1ES~1101-232 to have a power law described by ($\rm dN/dE \propto E^{-1.78 \pm 0.20_{stat}}$) and for the spectrum of 1ES~1218+304  to be as hard as ($\rm dN/dE \propto E^{-1.86 \pm 0.37_{stat}}$). All other EBL scenarios yield harder spectra for one of these three blazars.
A detailed analysis of the VHE spectra of 1ES~0229+200 was carried out by \citet{Aharonian:2007p1477}. This analysis supported an EBL spectrum and density close to the lower limits from the Spitzer measurements, and a hard intrinsic spectrum for the
blazar.
Figure \ref{fig6} shows the measured spectrum of  1ES~1218+304 and two possible intrinsic blazar spectra that are compatible with the limits from galaxy counts.

These results clearly indicate that blazar spectra, as evidenced by two of the most distant blazars (1ES~1101-232 and 1ES~1218+304), are  hard.  These values are still within the acceptable range predicted for shock acceleration in blazars (see, e.g., \citet{Stecker:2007p2209}), however they are also close to a limit ($\rm dN/dE \propto E^{-1.5} $) that was previously suggested by \citet{Aharonian:2006p1822}. The hard intrinsic spectrum of 1ES~1218+304 indicates that the peak in the VHE power output is located beyond $\sim 2$ TeV.
A high-energy peak above 2 TeV in the blazar spectral energy distribution was similarly seen by H.E.S.S. for both 1ES~1101-232
\citep{Aharonian:2007p2153} and 1ES~0229+200 \citep{Aharonian:2007p1477}.

In conclusion, results presented here from VERITAS observations confirm with high statistical significance the MAGIC discovery \citep{Albert:2006yq} of the HBL object 1ES 1218+304 as a source of VHE gamma rays. The normalization of the flux and spectral index both agree within errors with the MAGIC results. New limits on the density of the EBL in the near-IR to mid-IR could not be established. However, based on lower EBL limits from galaxy counts we were able to limit the range of intrinsic energy spectra for 1ES~1218+304 and showed that the intrinsic spectrum is harder than a power law with  $\rm  dN/dE \propto E^{-2.32 \pm 0.37_{stat}}$. When including constraints from the spectra of 1ES~1101-232 and 1ES~0229+200, the spectrum of 1ES 1218+304 is inferred to be harder than $\rm dN/dE \propto E^{-1.86 \pm 0.37_{stat}}$. 
Future deep observations at large zenith angles (where the effective area is larger for multi-TeV gamma-rays) could help extend the spectrum to 10 TeV and help constrain the EBL.
We see no evidence of a high energy peak in the SED for 1ES 1218+304 up to $\sim 2$ TeV; simultaneous, broadband measurements of the SED are required for detailed blazar modelling studies and to distinguish between leptonic and hadronic blazar models. 

\subsection*{Acknowledgments}

This research is supported by grants from the U.S. Department of Energy, the U.S. National Science Foundation, and the Smithsonian Institution, by NSERC in Canada, by PPARC in the UK and Science Foundation Ireland.




\clearpage
\input{tab1}

\clearpage




\clearpage








\end{document}

%% file: tab1.tex
\begin{deluxetable}{ccccccc}
\footnotesize
\tablecaption{The absorption-corrected spectral indices ($\rm dN/dE \propto E^{- \Gamma_{source}}$) of 1ES~1101-232 and 1ES~1218+304 and 1ES~0229+200 from a small sample of all EBL scenarios considered. EBL scenarios that fall below the lower limits from galaxy counts are marked with an asterisk.  Scaled Aharonian scenarios (AHA$x.y$)  are taken from \citet{Aharonian:2006p1822}. The errors represent only the statistical uncertainty.
\label{tbl-1}}
\tablewidth{0pt}
\tablehead{
\colhead{Scenario} & \multicolumn{3}{c}{$\Gamma_{source}$} \\
\cline{2-4}
 & \colhead{1ES~1101-232} &
\colhead{1ES~1218+304} &\colhead{ 1ES~0229+200}   }
\startdata
  AHA1.0      &   $\rm  0.44 \pm 0.20  $  &  $ \rm  0.35 \pm 0.40   $  & $ \rm  2.35 \pm 0.14$ \nl
  AHA0.55      &   $\rm 1.54 \pm 0.20  $  &  $ \rm  1.59 \pm 0.37   $   & $ \rm  2.41 \pm 0.14$  \nl
  AHA0.45      &   $\rm 1.78 \pm 0.20  $  &  $ \rm  1.86 \pm 0.37   $   & $ \rm  2.43 \pm 0.13   $   \nl
  HHH       &   $\rm -0.67 \pm 0.12  $  &  $ \rm  -0.72 \pm 0.29   $ & $ \rm  0.90 \pm 0.17   $  \nl
  LLH$^{*}$       &   $\rm 2.01 \pm 0.22  $  &  $ \rm  2.07 \pm 0.35   $  & $ \rm  2.12 \pm 0.20   $    \nl 
  LHL       &   $\rm 2.04 \pm 0.20  $  &  $ \rm  2.08 \pm 0.39   $  & $ \rm  0.94 \pm 0.32 $     \nl
  LHL0.70$^{*}$   &   $\rm 2.32 \pm 0.21  $  &  $ \rm  2.43 \pm 0.37   $  & $ \rm  1.43 \pm 0.29   $      \nl  
  LHL0.76   &   $\rm 2.23 \pm 0.21  $  &  $ \rm  2.32 \pm 0.37   $  & $ \rm  1.30 \pm 0.29   $      \nl
  LHL0.82   &   $\rm 2.18 \pm 0.21  $  &  $ \rm  2.26 \pm 0.38   $  & $ \rm  1.20 \pm 0.30   $      \nl
  MHL0.70  &   $\rm 1.26 \pm 0.19  $  &  $ \rm  1.34 \pm 0.36   $  &   $ \rm  1.35 \pm 0.21  $      \nl
  MHL0.55  &   $\rm 1.61 \pm 0.19  $  &  $ \rm  1.73 \pm 0.35   $  &   $ \rm  1.59 \pm 0.20  $     \nl
  LLL$^{*}$       &   $\rm 2.06 \pm 0.16  $  &  $ \rm  2.20 \pm 0.34   $    &  $ \rm  2.11 \pm 0.20   $      \nl
  SMS baseline &                       &  $  1.70 \pm 0.34$  &                \nl
\enddata
\end{deluxetable}

%% file: ms.bbl
\begin{thebibliography}

\bibitem[Aharonian et al.(2008a)]{Aharonian:2008p1531} Aharonian, F., et al. 2008, \aap, 478, 387

\bibitem[Aharonian et al.(2006)]{Aharonian:2006p1822} Aharonian, F., et al. 2006, Nature, 440, 1018

\bibitem[Aharonian et al.(2007a)]{Aharonian:2007p2153} Aharonian, F., et al. 2007a \aap, 470, 475

\bibitem[Aharonian et al.(2007b)]{Aharonian:2007p1477} Aharonian, F., et al. 2007b, \aap, 475, L9

\bibitem[Aharonian el al.(2007c)]{Aharonian:2007p762} Aharonian, F., et al. 2007c, \apj, 664, L71

\bibitem[Albert et al.(2006)]{Albert:2006yq} Albert, J., et al. 2006, \apjl, 642, L119

\bibitem[Albert et al.(2007)]{Albert:2007mz} Albert, J., et al. 2007, \apjl, 666, L17

\bibitem[Albert et al.(2008a)]{Albert:2007p3044} Albert, J. et al. 2008a, Phys. Lett. B 668, 253

\bibitem[Albert et al.(2008b)]{Albert:2008p3043} Albert, J., et al. 2008b, Nature, 320, 1752

\bibitem[Bade et al.(1998)]{Bade:1998ly} Bade, N., et al. 1998, \aap, 334, 459

\bibitem[Becker et al.(1995)]{Becker:1995p3168} Becker, R.H., et al. 1995, \apj, 450, 559

\bibitem[Berge et al.(2007)]{Berge:2007p689} Berge, D., et al. 2007, \aap, 466, 1219

\bibitem[B{\"o}ttcher(2007)]{Bottcher:2007fr} {B{\"o}ttcher}, M. 2007, \apss, 309, 95

\bibitem[Coppi \& Aharonian(1999)]{Coppi:1999p2590} Coppi, P.S., Aharonina, F.A. 1999, \apj, 521, L33

\bibitem[Costamante \& Ghisellini(2002)]{Costamante:2002rt} Costamante, L., \& Ghisellini, G. 2002, \aap, 384, 56 

\bibitem[Daniel et al.(2007)]{Daniel:2007lr} Daniel, M., et al. 2007, Proceedings of the 30th ICRC

\bibitem[Davies \& Cotton(1957)]{Davies:Cotton} Davies, J.M., Cotton, E.S. 1957, Solar Energy, 1, 16

\bibitem[Dwek \& Krennrich(2005)]{Dwek:2005ve} Dwek, E., \&  Krennrich, F. 2005, \apj, 618, 657

\bibitem[Fazio et al.(2004)]{Fazio2004} Fazio, G.G. et al. 2004, \apjs, 154, 39

\bibitem[Gaidos et al.(1996)]{Gaidos:1996p3096} Gaidos, J.A., et al. 1996, Nature, 383, 319

\bibitem[Gould \& Schr{\'e}der(1967)]{Gould:1967p2336} Gould, R.J. \& Schr{\'e}der, G.P. 1967, Phys. Rev., 155, 1404

\bibitem[Hartman et al.(1999)]{hartman} Hartman, R.C., et al. 1999, \apjs, 123, 79

\bibitem[Hauser \& Dwek(2001)]{Hauser:2001p2604} Hauser, M.G., Dwek, E. 2001, \araa, 39, 249

\bibitem[Helene(1983)]{helene} Helene, O. 1983, Nucl. Instr. and Method, 212, 319 

\bibitem[Hillas(1985)]{hillas85} Hillas, A.M. 1985, Proceedings of the 19th ICRC

\bibitem[Hofmann et al.(1999)]{Hofmann:1999lr} Hofmann, W., et al. 1999, Astroparticle Physics, 12, 135

\bibitem[Holder et al.(2006)]{Holder:2006p2560}Holder, J., et al. 2006, Astroparticle Physics, 25, 391

\bibitem[Konopelko et al.(1995)]{Konopelko} Konopelko, A. et al. 1995, Proc. of the Padova Workshop on TeV Gamma-Ray Astrophysics ``Towards a Major Atmospherics Cherenkov Detector-IV", (ed. M. Cresti), Padova, Italy, 373

\bibitem[Krawczynski et al.(2006)]{Krawczynski:2006lr} Krawczynski, H., et al. 2006, Astroparticle Physics, 25, 380

\bibitem[Li and Ma(1983)]{Li:1983lr} Li, T.-P. and Ma, Y.-Q. 1983, \apj, 272, 317

\bibitem[Madau \& Pozetti (2000)]{Madau2000} Madau, P., \& Pozetti 2000, \mnras, 312, L9

\bibitem[Maier et al.(2007)]{Maier:2007kx} Maier, G., et al. 2007, Proceedings of the 30th ICRC

\bibitem[Malkan \& Stecker(1998)]{Malkan:1998p2653} Malkan, M.A., Stecker, F.W. 1998, \apj, 496, 13

\bibitem[Mazin \& Raue(2007)]{Mazin:2007p2248} Mazin, D., Raue, M. 2007, \aap, 471, 439

\bibitem[Moderski et al.(2005)]{Moderski:2005p2601} Moderski, R., et al. 2005, \mnras, 363, 954

\bibitem[Padovani \& Giommi(1995)]{Padovani:1995p2580} Padovani, P., Giommi, P. 1995, \apj, 444, 567

\bibitem[Primack(2001)]{Primack:2002p2585} Primack, J.R. 2001, AIP Conf.Proc. 558, 463

\bibitem[Stecker et al.(1992)]{Stecker:1992p2571} Stecker, F.W., et al. 1992, \apj, 390. L49

\bibitem[Stecker et al.(2006)]{Stecker:2006p2224} Stecker, F.W., Malkan, M.A., Scully, S.T. 2006, \apj, 648, 774

\bibitem[Stecker et al.(2007)]{Stecker:2007p2209} Stecker, F.W., Barring, M.G., Summerlin, E.J. 2007, \apj 667, L29

\bibitem[Swordy (2008)]{Swordy:2008} Swordy, S. 2008, ATel 1422

\bibitem[Weekes et al.(2002)]{T.-C.-Weekes:2002lr} Weekes, T.~C., et al. 2002, Astroparticle Physics, 17, 221


\end{thebibliography}
